\documentclass[
 aps,           
pra,            
twocolumn,      
letterpaper,    
showpacs,       
preprintnumbers,
amsmath,        
amssymb,        
floatfix]{revtex4-2}
\usepackage[T1]{fontenc}
\usepackage[utf8]{inputenc}
\usepackage{ulem}
\usepackage{graphicx}    
\usepackage{color}      
\usepackage{bm}         
\usepackage{braket}     
         
\usepackage{amsmath,stfloats}
\usepackage{xcolor}
\usepackage{xpatch}
\usepackage{etoolbox}

\begin{document}
\title{\textbf{Geometric quantum discord in the black hole quantum atmosphere} } 

 \author{Siwei Li}
 \affiliation{School of Mathematics and Statistics, Hainan Normal University, Haikou 571158, China}
 \author{Xiaofen Huang} 
 \email{huangxf1206@163.com}
 \affiliation{School of Mathematics and Statistics, Hainan Normal University, Haikou 571158, China}
 
 \begin{abstract}
We investigate the geometric quantum discord of bipartite Werner states influenced by Hawking radiation in the quantum atmosphere of a Schwarzschild black hole. We find that the geometric quantum discord in the physically accessible region exhibits a nonmonotonic behavior: it first decreases and then increases as the normalized radial distance increases, while the discord in the physically inaccessible region shows the exact opposite trend. A pronounced extremum of geometric quantum discord occurs precisely at the position corresponding to the peak intensity of Hawking radiation. Furthermore, we find that the Hartle–Hawking constant and the event horizon radius exert opposite effects on the strength of quantum correlation redistribution: a larger Hartle–Hawking constant enhances the redistribution effect, whereas a larger event horizon radius suppresses it.
\end{abstract}

\maketitle

\section{Introduction}
Quantum discord serves as a measurement-based classification scheme for quantum correlations, which bears fundamental significance and broad application prospects in quantum information science     \cite{luo2008using,li2008classical,radhakrishnan2020multipartite,inui2020entanglement,zhou2020quantum}. Mathematically, it refers to the minimum gap between two quantum formulations of classically equivalent mutual information obtained under projective measurement operations\cite{luo2010geometric,rulli2011global}. For a bipartite quantum state $\rho$ on quantum system $\mathcal{H}_A \otimes \mathcal{H}_B$, where $A$ and $B$ are the individual subsystems, the quantum discord is defined as follows,
\begin{equation}
	D(\rho) = I(\rho) - \min_{\Pi^A} I\left[\Pi^A(\rho)\right],
	\label{eq:discord_def}
\end{equation}
here the minimum is over von Neumann measurements (one dimensional orthogonal projectors summing to the identity) $\Pi^A = \{\Pi_k^A\}$  on party $A$, and
\[
\Pi^A(\rho) = \sum_k \left(\Pi_k^A \otimes I^B\right) \rho \left(\Pi_k^A \otimes I^B\right)
\]
is the resulting state after the measurement. $I(\rho) = S(\rho_A) + S(\rho_B) - S(\rho)$ is the quantum mutual information, $S(\rho) = -\operatorname{Tr}\rho \log_2 \rho$ is the von Neumann entropy. The intuitive meaning of quantum discord may be interpreted as the minimal loss of correlations (as measured by the quantum mutual information) due to measurement. This formulation of quantum discord is equivalent to the original definition of quantum discord by Ollivier and Zurek \cite{ollivier2001quantum}. 

Hawking radiation \cite{Damour:1976:PRD,hawking1975cmp,hawking1976prd} established the first fundamental connection between general relativity and quantum mechanics, marking a milestone in modern theoretical physics. Understanding the evolution of quantum correlations in curved spacetimes is therefore essential for reconciling these two foundational theories and addressing the long-standing black hole information paradox \cite{bombelli1986}. The conventional picture holds that Hawking radiation stems from quantum excitations in the near-horizon region with $\Delta r = r - r_h \ll r_h$, where $r_h$ is the event horizon radius \cite{hawking1976prd,hawking1975cmp,unruh1977prd}. Nevertheless, through analyses of the total emission rate and stress tensor associated with Hawking radiation, Giddings demonstrated that the radiation originates from the near-horizon quantum region, also referred to as the ``quantum atmosphere''. Its radial range is scaled by the horizon radius, i.e., $\Delta r = r_A - r_h \sim r_h$ \cite{giddings2016hawking}. This revised understanding of the spatial distribution of Hawking radiation requires that we reexamine quantum correlation dynamics in this extended region.

Recent work in relativistic quantum information has focused on non-inertial spacetimes, with extensive studies investigating the behavior of quantum correlations, including entanglement \cite{zhang2025entanglement,liu2025entanglement,liu2023fermionic,wang2020genuine,mi2024impact}, coherence \cite{Kaczmarek2025Coherence,li2025multiqubit,wu2021quantum,liao2025quantum}, nonlocality \cite{Kaczmarek2024Signatures,Kaczmarek2026Nonlocal,
	mi2025genuine,zhang2023hawking}, uncertainty relations \cite{zhang2025entanglement,Wang2024Entropic},  quantum steering \cite{wu2025fermionic,wu2025gaussian,liu2018influence}, and so on. Key results show these correlations are degraded by information loss due to Hawking radiation \cite{liu2018satellite,liu2019influence,wang2011multipartite,torres2019entanglement,wu2020quantum}, findings that advance both our understanding of quantum information in curved backgrounds and the study of the black hole information paradox and entanglement entropy. Despite these significant advances, the local dynamics inside the quantum atmosphere remain largely unexplored, and how key physical parameters modulate the redistribution of quantum correlations has yet to be systematically elucidated. Compared with quantum entanglement, nonlocality, and quantum coherence, quantum discord covers a broader spectrum of nonclassical correlations, including quantum correlations in separable states. Unlike entanglement, quantum discord can persist in separable states and exhibits greater robustness against environmental decoherence, making it a more general and practically relevant measure for studying quantum information in noisy environments such as black hole spacetimes. As a continuous measure, quantum discord is capable of finely characterizing the nonmonotonic behavior of correlation redistribution near the peak of the local temperature, a feature that Bell-type binary measures cannot achieve. Moreover, geometric quantum discord enjoys the computational advantage of a closed-form analytical expressionquantum discord \cite{huang2026}, allowing it to be rigorously solved in curved spacetime frameworks without resorting to complicated extremal optimization.

In this work, we focus on the investigation of the dynamics of geometric quantum discord  in the quantum atmosphere of a Schwarzschild black hole. We adopt the local Hartle–Hawking temperature profile to accurately describe the Hawking effect within the quantum atmosphere. Specifically, we derive analytical expressions for the geometric quantum discords in both physically accessible and inaccessible regions, and establish  complementary trade-off relations between the quantum discords in these two regions. Most notably, we find that the extremal values of quantum discord precisely correspond to the peak intensity of Hawking radiation in the quantum atmosphere. Furthermore, we systematically analyze the effects of the normalized radial distance $r/r_h$, event horizon radius $r_h$, and Hartle–Hawking constant $D_{HH}$ on the quantum discord redistribution process, and reveal that $r_h$ and $D_{HH}$ exert opposite regulatory effects on this process. These findings provide new insights into exploring the information structure of black hole quantum atmospheres and the quantum nature of Hawking radiation. 

The remainder of this paper is organized as follows. In Section II, we introduce the geometric measure of quantum discord and derive its analytical expressions for  Werner states under the influence of the Schwarzschild black hole quantum atmosphere. In Section III, we discuss in detail the evolution behaviors of geometric quantum discords in both physically accessible and inaccessible regions under the influence of the black hole quantum atmosphere, and systematically characterize their dependencies on the key physical parameters. At last, section IV gives a brief summary of the whole work.

\section{Geometric quantum discord in the black hole quantum atmosphere}
Generally, the metric of the Schwarzschild black hole is denoted by
\begin{equation}
	\begin{split}
		ds^2 &= -\left(1-\frac{2M}{r}\right)dt^2 
		+ \left(1-\frac{2M}{r}\right)^{-1}dr^2 \\
		&\quad + r^2\left(d\theta^2+\sin^2\theta\,d\varphi^2\right),
	\end{split}
\end{equation}
where $M$ denotes the black hole mass \cite{Martinez:2010:entanglement_blackhole}, $r$ is the radial coordinate, $t$ denotes the time coordinate, $\theta$ denotes the polar angle, and $\varphi$ denotes the azimuthal angle. Throughout this work, we adopt natural units where $G=c=\hbar=k_B=1$ for simplicity. In Schwarzschild spacetime, the curved-spacetime Dirac equation $\left[\gamma^a e^\mu_a\left(\partial_\mu+\Gamma_\mu\right)\right]\psi=0$. Substituting the Schwarzschild vierbein into the Dirac equation, we obtain the explicit form of the massless Dirac equation in Schwarzschild spacetime \cite{Jing:2004:late_time_dirac}: 
\begin{equation}
	\begin{split}
		&\gamma_1\sqrt{1-\frac{2M}{r}}\left[\frac{\partial}{\partial r}+\frac{1}{r}+\frac{M}{2r(r-2M)}\right]\psi-\frac{\gamma_0}{\sqrt{1-\frac{2M}{r}}}\frac{\partial\psi}{\partial t} \\
		&+\frac{\gamma_2}{r}\left(\frac{\partial}{\partial\theta}+\frac{\cot\theta}{2}\right)\psi + \frac{\gamma_3}{r\sin\theta}\frac{\partial\psi}{\partial\varphi}=0,
	\end{split}
\end{equation}
where $\gamma_i$ ($i = 0, 1, 2, 3$) represent Dirac gamma matrices. 

By solving the above equation, two sets of positive-energy fermionic solutions are obtained:
\begin{equation}
	\psi_k^{\mathrm{I}+}=\xi e^{-\mathrm{i}\omega u}, \quad \psi_k^{\mathrm{II}+}=\xi e^{\mathrm{i}\omega u},
\end{equation}
where $\omega$ is the monochromatic frequency of the Dirac field, $k$ represents the field mode, $\xi$ is the 4-component Dirac spinor,  $u = t - r_*$, 
and the tortoise coordinate is defined as
\begin{equation}
	r^* = r + 2M \ln\left( \frac{r}{2M} - 1 \right).
\end{equation}

The positive frequency solutions of Eq.~(4) correspond to the interior region and the exterior region of the event horizon, respectively. Using the Damour and Ruffini method \cite{Damour:1976:PRD}, the positive energy mode (Kruskal mode) is used to connect the two equations of Eq.~(4)
\begin{equation}
	\begin{aligned}
		\Phi_{k,\mathrm{I}}^+ &= e^{-2\pi M\omega}\Psi_{-k,\mathrm{II}}^- + e^{2\pi M\omega}\Psi_{k,\mathrm{I}}^+, \\
		\Phi_{k,\mathrm{II}}^+ &= e^{-2\pi M\omega}\Psi_{-k,\mathrm{I}}^- + e^{2\pi M\omega}\Psi_{k,\mathrm{II}}^+.
	\end{aligned}
	\label{eq:bogoliubov_kruskal}
\end{equation}

In Kruskal coordinates, the Dirac field $\psi$ can be expanded as:
\begin{equation}
	\begin{aligned}
		\psi &= \int dk\left[2\cosh(4\pi M\omega)\right]^{-\frac{1}{2}}\left[\hat{c}_k^{\mathrm{II}}\Psi_{k,\mathrm{II}}^+ + \hat{d}_{-k}^{\mathrm{II}\dagger}\Psi_{-k,\mathrm{II}}^- \right. \\
		&\left.+ \hat{c}_k^{\mathrm{I}}\Psi_{k,\mathrm{I}}^+ + \hat{d}_{-k}^{\mathrm{I}\dagger}\Psi_{-k,\mathrm{I}}^-\right],
	\end{aligned}
\end{equation}
where $\hat{c}_k$ and $\hat{d}^\dagger_{-k}$ are the annihilation and creation operators of the Kruskal vacuum state.

From the Bogoliubov transformation, the Kruskal vacuum and excited states in the Schwarzschild spacetime can be expressed as:
\begin{equation}
	\begin{aligned}
		|0\rangle_k &= \frac{1}{\sqrt{e^{-\frac{\omega}{T_H}}+1}}|0\rangle_\mathrm{I}|0\rangle_\mathrm{II} + \frac{1}{\sqrt{e^{\frac{\omega}{T_H}}+1}}|1\rangle_\mathrm{I}|1\rangle_\mathrm{II}, \\
		|1\rangle_k &= |1\rangle_\mathrm{I}|0\rangle_\mathrm{II},\label{eq:transform}
	\end{aligned}
\end{equation}
where $T_H = \frac{1}{8\pi M}$ denotes the Hawking temperature \cite{Damour:1976:PRD}, $\omega$ is the particle frequency, and $\{|n\rangle_{\mathrm{I(II)}}\}$ indicate the fermionic (antifermionic) modes in Region $\mathrm{I}$ ($\mathrm{II}$). To simplify the subsequent analysis, we define the substitution $\frac{1}{\sqrt{e^{-\frac{\omega}{T_H}} + 1}} = \cos\alpha$ and $\frac{1}{\sqrt{e^{\frac{\omega}{T_H}} + 1}} = \sin\alpha$, $\omega = 1$, where $\alpha$ denotes the parameter.

Next, we fucus on considering the quantum discord of mixed states in black hole quantum atmosphere. Quantum discord as a measure of quantum correlations, initially introduced by Ollivier and Zurek \cite{ollivier2001quantum} and by Henderson and Vedral \cite{henderson2001classical}, is attracting increasing interest  \cite{luo2008quantum,zhu2021geometric,li2021quantum,hunt2019how,szasz2019measure,zhou2025analytic}. Recently Daki\'c et al \cite{dakic2010necessary}, proposed the following geometric measure of quantum discord for bipartite states:
\begin{equation}
	D_\mathrm{G}(\rho) = \min_{\chi \in \epsilon} \|\rho - \chi\|,
\end{equation}
where $\epsilon$ denotes the set of zero-discord states and the geometric quantity $\|\rho - \chi\|^2 = \mathrm{Tr}(\rho - \chi)^2$ is the square of the Hilbert--Schmidt norm of Hermitian operators.

Since the Pauli operators $\{\sigma_j\}$, where $j = 0, 1, 2, 3$, form an orthogonal basis for the qubit system, for any two-qubit state $\rho$, it can be decomposed into a linear combination based on the operator bases $\{\sigma_j\}$, that is the Bloch representation as follows\textcolor{red}{,}

\begin{equation}
	\begin{split}
		\rho = \frac{1}{4}\bigg(
		&I^A \otimes I^B 
		+ \sum_{i=1}^{3} x_i \sigma_i \otimes I^B 
		+ \sum_{j=1}^{3} y_j I^A \otimes \sigma_j \\
		&+ \sum_{i,j=1}^{3} g_{ij} \sigma_i \otimes \sigma_j
		\bigg),
	\end{split}
\end{equation}
where $I^A$ ($I^B$) being the identity matrices of subsystem $A$ ($B$) and $x_i = \mathrm{Tr}\left[\rho(\sigma_i \otimes I^B)\right]$, $y_j = \mathrm{Tr}\left[\rho(I^A \otimes \sigma_j)\right]$, $g_{ij} = \mathrm{Tr}\left[\rho(\sigma_i \otimes \sigma_j)\right]$ denoting the correlation coefficients.

Therefore, the geometric measure of quantum discord of a two-qubit state $\rho$ is evaluated as \cite{dakic2010necessary}
 \begin{equation}
	D(\rho) = \frac{1}{4}\left(\|\mathbf{x}\|^2 + \|G\|^2 - \lambda_{\max}\right),\label{eq:quantum discord}
\end{equation}
	where $\mathbf{x} := (x_1, x_2, x_3)^t$ is named the Bloch vector, its length is defined by 
	$\|\mathbf{x}\| := \sqrt{\sum_i x_i^2}$, while $G := (g_{ij})$ is the correlation matrix with size $3\times 3$, and $\lambda_{\max}$ is the largest eigenvalue of the matrix $\mathbf{x}\mathbf{x}^t + G G^t$. Here the superscript $t$ denotes transpose of vectors or matrices. Let two observers Alice and Bob share the following Werner state \cite{Werner:1989},
	\begin{equation}
		\rho_W = \frac{1}{6}\left[(2-p)I^A \otimes I^B + (2p-1)F\right],\label{eq:werner sate}
	\end{equation}
	where state parameter $-1 \leq p \leq 1$ and $F$ is the ``flip'' or ``swap'' operator defined by $F(\phi \otimes \psi) = \psi \otimes \phi$. Furthermore, Werner state (\ref{eq:werner sate}) has Bloch representation
	\begin{equation}
		\rho_W = \frac{1}{4}\left(
		I^A \otimes I^B + \frac{2p-1}{3}\sum_{i=1}^3 \sigma_i \otimes \sigma_i
		\right).	
	\end{equation}
	
	At first, both Alice and Bob reside in the asymptotically flat region. Thereafter, Alice remains in this region, while Bob undergoes free infall into the Schwarzschild black hole. Therefore, Bob will be subject to the effects of Hawking radiation in the quantum atmosphere \cite{giddings2016hawking} beyond the event horizon radius $r_h$. So the Werner state $\rho_W$ will be transformed to a tripartite quantum state $\rho_{A B_\text{I} B_\text{II}}$, whose analytical form can be derived from Eqs.~(\ref{eq:transform}) which is omitted here for its complicated structure. Given the disconnection between regions I and II, and the fact that Bob cannot access the modes within the event horizon, mode $B_\mathrm{I}$ located outside the event horizon is referred to as physically accessible, whereas mode $B_\mathrm{II}$ situated inside the event horizon is  designated as inaccessible. Through tracing over all degrees of freedom in region II, the state of the physically accessible subsystem I can be captured, thereby leading to the reduced density matrix $\rho_{A B_\mathrm{I}}$, with (detailed calculations are presented in the Appendix),
\begin{equation}
	\begin{split}
		\rho_{AB_\mathrm{I}} &= \frac{1}{4}\bigg(
		I^A \otimes I^B 
		- I^A \otimes \sin^2\alpha\,\sigma_3 
		+ \frac{2p-1}{3}\cos\alpha\,\sigma_1 \otimes \sigma_1 \\
		&\quad + \frac{2p-1}{3}\cos\alpha\,\sigma_2 \otimes \sigma_2
		+ \frac{2p-1}{3}\cos^2\alpha\,\sigma_3 \otimes \sigma_3
		\bigg).
	\end{split}
\end{equation}
After tracing out all degrees of freedom in region I, the quantum state of the physically inaccessible subsystem II is obtained, from which we can derive the reduced density matrix $\rho_{AB_\mathrm{II}}$ with,
\begin{align}
	\begin{split}
		\rho_{AB_\mathrm{II}} &= \frac{1}{4}\bigg(
		I^A\otimes I^B + I^A\otimes \cos^2\alpha\,\sigma_3 +\frac{2p-1}{3}\sin\alpha\,\sigma_1\otimes\sigma_1 \\ 
		&\quad- \frac{2p-1}{3}\sin\alpha\,\sigma_2\otimes\sigma_2
		- \frac{2p-1}{3}\sin^2\alpha\,\sigma_3\otimes\sigma_3
		\bigg).
	\end{split}
\end{align}
    
Employing Eq.~(\ref{eq:quantum discord}), we obtain the discord of reduce quantum state $\rho_{A B_\mathrm{I}}$ as follows,	
\begin{equation}
	D(\rho_{AB_\mathrm{I}}) = \frac{(2p-1)^2}{36} \cdot \frac{e^{\frac{\omega}{T}} \left( 2e^{\frac{\omega}{T}} + 1 \right)}{\left( e^{\frac{\omega}{T}} + 1 \right)^2}.
    \label{16}
\end{equation}

In the similarly way, we calculate the another reduce density matrices and the corresponding discords (see detailed calculations in the Appendix), their analytical expressions can be written out explicitly in the following,
\begin{equation}
	D(\rho_{AB_\mathrm{II}}) = \frac{(2p-1)^2}{36} \cdot \frac{e^{\frac{\omega}{T}} + 2}{\left(e^{\frac{\omega}{T}} + 1\right)^2}.
    \label{17}
\end{equation}

We now consider the scenario where both Alice and Bob freely fall into a Schwarzschild black hole, the Werner state $\rho_W$ will be transformed to a four-partite quantum state $\rho_{A_\text{I}A_\text{II}B_\text{I}B_\text{II}}$, {whose analytical expression can be
	obtained using Eq.~(8) since the interior and exterior regions of the event horizon are causally disconnected, we refer to the modes inside the event horizon, i.e., $A_{\text{II}}$ and $B_{\text{II}}$, as physically inaccessible modes, and the modes outside the event horizon ($A_{\text{I}}$ and $B_{\text{I}}$) as physically accessible modes. As Alice and Bob have no access to the interior of the event horizon, we trace out the modes $A_{\text{II}}$ and $B_{\text{II}}$ to obtain the bipartite reduced density matrix $\rho_{A_{\text{I}} B_{\text{I}}}$.
	\begin{equation}
		\begin{split}
			\rho_{A_\mathrm{I} B_\mathrm{I}}
			&= \frac{1}{4}\Bigl(
			I^A \otimes I^B
			- \sin^2\alpha \, \sigma_3 \otimes I^B
			- \sin^2\alpha \, I^A \otimes \sigma_3
			\\
			&\quad
			+ \frac{2p-1}{3}\cos^2\alpha \, \sigma_1 \otimes \sigma_1
			\\
			&\quad
			+ \frac{2p-1}{3}\cos^2\alpha \, \sigma_2 \otimes \sigma_2
			\\
			&\quad
			+ \Bigl[\sin^4\alpha + \frac{2p-1}{3}\cos^4\alpha\Bigr]
			\sigma_3 \otimes \sigma_3
			\Bigr).
		\end{split}
		\end{equation}
Employing Eq.~(11), we obtain the discord of reduce quantum state $\rho_{A_{\text{I}} B_{\text{I}}}$ as follows,
	\begin{align}
	D(\rho_{A_\text{I} B_\text{I}}) &= \frac{(2p-1)^2}{18} \cdot \frac{e^{\frac{2\omega}{T}}}{\left(e^{\frac{\omega}{T}} + 1\right)^2}.
    \label{19}
	\end{align}
Similarly, we calculate the other five reduced density matrices and the associated discords, whose explicit analytical forms are given below:
\begin{align}
	D(\rho_{A_\text{II} B_\text{II}}) &= \frac{(2p-1)^2}{18} \cdot \frac{1}{\left(e^{\frac{\omega}{T}} + 1\right)^2}, \\
D(\rho_{A_\text{I} A_\text{II}}) &= D(\rho_{B_\text{I} B_\text{II}}) = \frac{1}{4} \cdot \frac{1}{(e^{\frac{\omega}{T}} + 1)}, \\   
	D(\rho_{A_\text{I} B_\text{II}}) &= D(\rho_{A_\text{II} B_\text{I}}) = \frac{(2p-1)^2}{18} \cdot \frac{e^{\frac{\omega}{T}}}{\left(e^{\frac{\omega}{T}} + 1\right)^2}.
\end{align}

It is worth noting that quantum discords $D(\rho_{A_\text{I}A_\text{II}})$, $D(\rho_{B_\text{I}B_\text{II}})$ are independent of the state parameter $p$. 
As \(e^{\frac{\omega}{T}}>1\), discords \(D(\rho_{A_\text{I} B_\text{II}}) \) and \(D(\rho_{A_\text{II} B_\text{I}})\) admit an upper bound below,
\begin{equation}
D(\rho_{A_\text{I} B_\text{II}}) = D(\rho_{A_\text{II} B_\text{I}}) < \frac{(2p-1)^2}{72} \leq \frac{1}{8}.
\end{equation}

Furthermore, it can be easily checked that there is a trade-off relation among geometric measure of quantum discords $D(\rho_{A_\text{I}B_\text{I}})$, $D(\rho_{A_\text{II}B_\text{I}})$, $D(\rho_{A_\text{I}B_\text{II}})$ and $D(\rho_{A_\text{II}B_\text{II}})$, that is,
\begin{equation}
D(\rho_{A_\text{I} B_\text{I}}) \cdot D(\rho_{A_\text{II} B_\text{II}}) = D(\rho_{A_\text{I} B_\text{II}}) \cdot D(\rho_{A_\text{II} B_\text{I}}).
\end{equation}

This conservation of the relative ratio indicates that Hawking radiation distributes quantum discords unevenly between the physically accessible and inaccessible regions. 

\section{thermal evolution of quantum discord}

To investigate the potential characteristics of the black hole's quantum atmosphere, we can replace the Hawking temperature $T$ in Eqs.~\eqref{16}, \eqref{17}, \eqref{19} and (20) to (22) with the local temperature $T_{HH}$ in Hartle-Hawking vacuum \cite{eune2019}, which can be expressed as
\begin{equation}
		\begin{split}
			&T_{HH} = T_H \sqrt{1 - \frac{r_h}{r}}\\ 
			&\sqrt{1 + 2\frac{r_h}{r} + \left(\frac{r_h}{r}\right)^2 \left(9 + 4D_\text{HH} + 36\ln\left(\frac{r_h}{r}\right)\right)},
		\end{split}
\end{equation}
where $T_{H} = \dfrac{1}{4\pi r_{h}}$, $D_{\text{HH}}$ is the undetermined constant of the stress tensor in the Hartle–Hawking vacuum, namely the Hartle–Hawking parameter. 
The constant $D_\text{HH}$ in Eq.~(25) is considered arbitrary. The Hartle--Hawking boundary conditions alone are insufficient to fix $D_\text{HH}$, and thus additional conditions are required to determine its value. To avoid an imaginary temperature in the region outside the horizon and unphysical inverse scaling of temperature with distance, the temperature remains real throughout the entire region for $D_\text{HH} \geq D_{\text{c}} \simeq 23.03$.
In the following discussion, we impose the condition $D_{\text{HH}} \geq D_{\text{c}}$, $r_{\text{h}} > 0$. As $D_{\text{HH}}$ increases, the peaks of local equilibrium temperatures satisfy $1.43\,r_{\text{h}} \lesssim r_{\text{peak}} < 1.5\,r_{\text{h}}$ \cite{eune2019}. It is also worth noting that $T_{HH} = 0$ at the event horizon of black hole, and asymptotically approaches the standard Hawking temperature of the Schwarzschild black hole as $r\to\infty$.

To gain an in-depth insight into the close relationship between discord and quantum atmosphere, we treat
the physically accessible and physically inaccessible quantum discords $D(\rho_{AB_\mathrm{I}})$ and $D(\rho_{AB_{\mathrm{II}}})$ as functions of the normalized distance $r/r_h$ for different values of the constant $\mathrm{D_{HH}}$ with state parameter $p = \sqrt{10/11}$. 

As shown in Fig.~\ref{fig1}(a), it is interesting to ﬁnd that: (i) The physically accessible quantum discord $D(\rho_{AB_\mathrm{I}})$ first decreases and then increases as the normalized distance $r/r_h$ increases, and it eventually converges to a value of about 0.0457, which corresponds to the maximum quantum discord. (ii) The extreme values of quantum discord correspond to the peaks of Hawking radiation detected outside the event horizon, which occur in the interval $r/r_h \in [1.43, 1.5)$. Furthermore, the larger the value of $\mathrm{D_{HH}}$, the greater the quantum discord loss. For all values of $\mathrm{D_{HH}}$, the minimum quantum discord always occurs in the interval $1.43 \lesssim r/r_h < 1.5$.

By contrast, the quantum discord $D(\rho_{AB_\mathrm{II}})$ in the physically inaccessible region first increases and then decreases as $r/r_h$ increases, which is nearly inverse to the quantum discord in the physically accessible region, as shown in Fig.~\ref{fig1}(b). This can be interpreted by the fact that the Hawking effect associated with black holes induces a redistribution of quantum discord in the current setup. Specifically, a stronger Hawking effect diminishes bipartite quantum discord in the physically accessible region while enhancing it in the physically inaccessible region.
\begin{figure}
	\centering
	\includegraphics[width=1\linewidth]{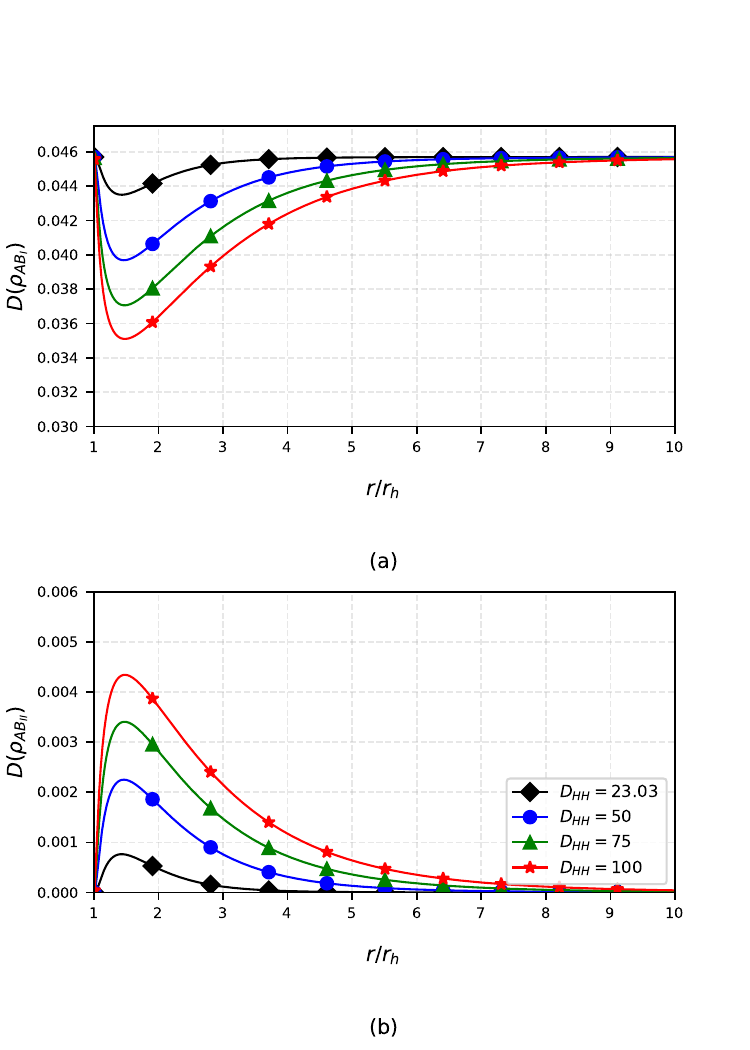}
	\caption{\label{fig1}Quantum discord of the quantum systems as functions of the normalized distance $r/r_h$ for different constant values of $\mathrm{D_{HH}}$. 
		Panel (a): Physically accessible quantum discord D($\rho_{AB_\mathrm{I}}$); 
		Panel (b): Physically inaccessible quantum discord D($\rho_{AB_\mathrm{II}}$). 
		The state parameter p is fixed at $p = \sqrt{10/11}$.}
\end{figure}

 In Fig.~\ref{fig2} and Fig. \ref{fig3}, it is interesting to ﬁnd that: (i) For any fixed state parameter $p$, the physically accessible quantum discord $D(\rho_{AB_\mathrm{I}})$ first decreases and then increases, which is consistent with the phenomenon presented in Fig.~\ref{fig1}. (ii) In contrast,  for any fixed state parameter $p$, the physically inaccessible quantum discord $D(\rho_{AB_\mathrm{II}})$ increases first and then decreases. (iii) As the state parameter $p$ increases, the quantum discord first decreases and then increases, reaching its global minimum at $p=1/2$. Furthermore, the quantum discord curve is symmetric about $p=1/2$. (iv) A stronger initial-state quantum discord (corresponding to values of $p$ farther from $1/2$) ensures that a non-negligible amount of quantum discord persists in the physically accessible region even where Hawking radiation is strongest, which demonstrates that strongly correlated initial states possess superior robustness to Hawking-induced decoherence.
\begin{figure}
	\centering
	\includegraphics[width=1\linewidth,clip,trim={2cm 0 0 0}]{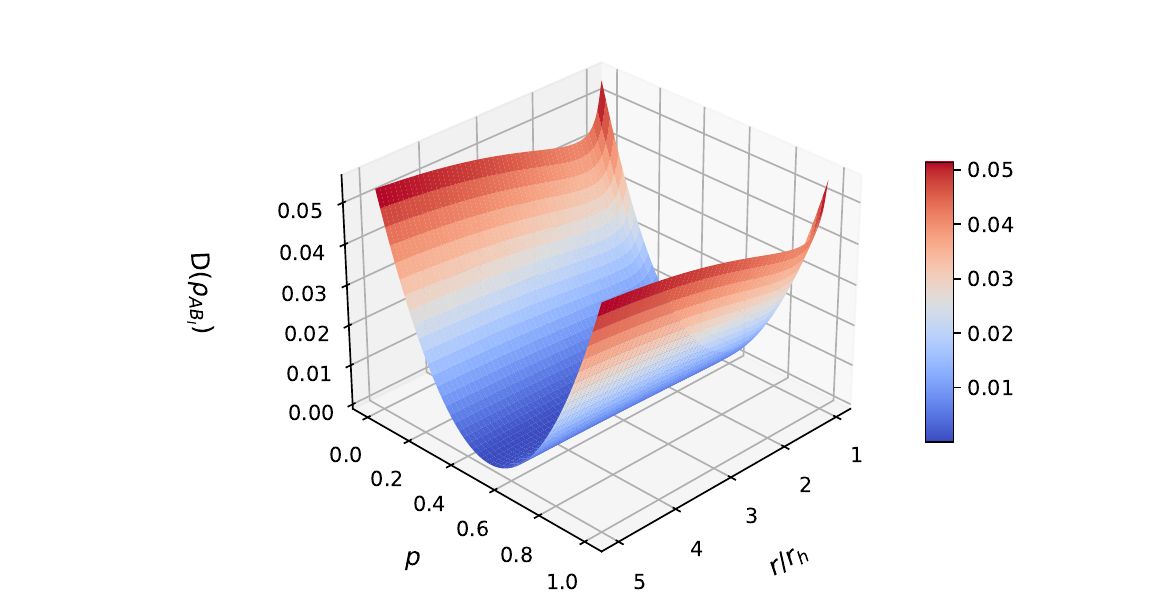}
	\caption{\label{fig2}Physically accessible quantum discord D($\rho_{AB_\mathrm{I}}$) as functions of the normalized distance $r/r_h$ and the state parameter $p$ at $\mathrm{D_{HH}} = 100$.}
\end{figure}
\begin{figure}
	\centering
	\includegraphics[width=1\linewidth,clip,trim={2cm 0 0 0}]{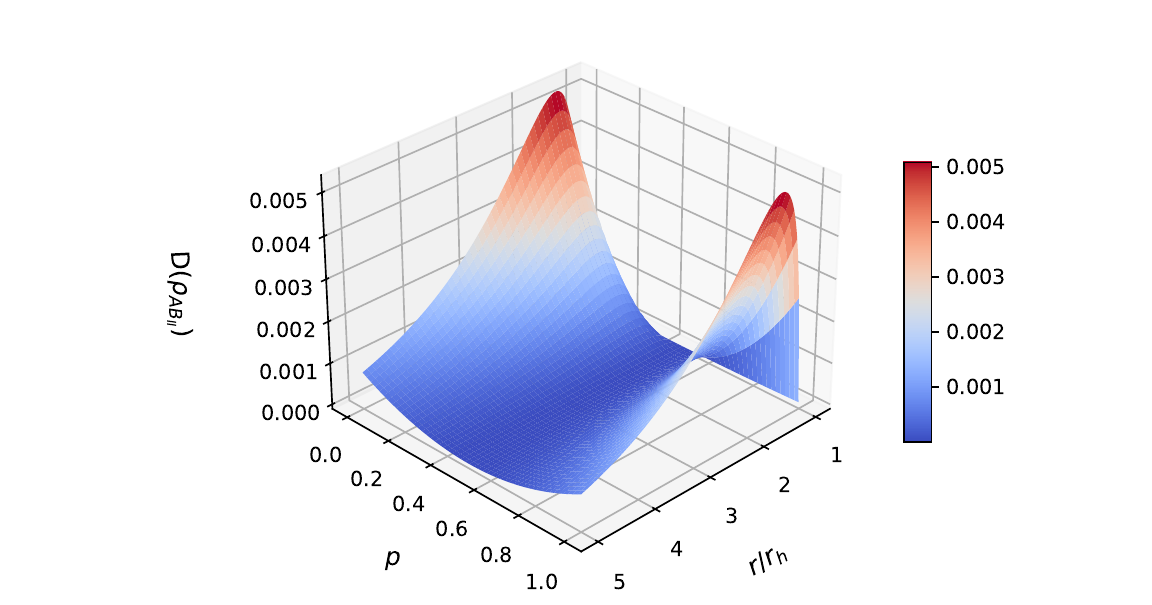}
	\caption{\label{fig3}Physically inaccessible quantum discord D($\rho_{AB_\mathrm{II}}$) as functions of the normalized distance $r/r_h$ and the state parameter $p$ at $\mathrm{D_{HH}} = 100$.}
\end{figure}

To investigate the potential correlation between $r_h$, $r$ and $\mathrm{D_{HH}}$, we plot Figs. \ref{fig4} and Fig. \ref{fig5}.
 From these ﬁgures, we obtain that: (i) Overall, the quantum discord in the physically accessible region first decreases and then increases with increasing $r$, while that in the physically inaccessible region exhibits the exact opposite trend. (ii) With increasing event horizon radiu $r_h$, the redistribution effect of quantum discord between the physically accessible and inaccessible regions is weakened.
(iii) With increasing $\mathrm{D_{HH}}$, the redistribution effect of quantum discord between the physically accessible and inaccessible regions is enhanced.

In Fig.~\ref{fig6}, we plot the geometric measure of quantum discord $D(\rho_{A_\text{I}B_\text{I}})$, $D(\rho_{A_\text{I}B_\text{II}})$,
$D(\rho_{A_\text{II}B_\text{II}})$ and $D(\rho_{A_\text{I}A_\text{II}})$ as functions of the normalized distance $r/r_h$. These plots reveal an interesting phenomenon: as the normalized distance $r/r_h$ increases, $D(\rho_{A_\text{I}B_\text{I}})$ first decreases and then increases, approaching approximately 0.0457, while $D(\rho_{A_\text{I}B_\text{II}})$, $D(\rho_{A_\text{II}B_\text{II}})$ and $D(\rho_{A_\text{I}A_\text{II}})$ first increase and then decrease, eventually converging to zero. This phenomenon indicates that Hawking radiation suppresses the quantum discord in the physically accessible region while simultaneously generating quantum discord in the physically inaccessible region. However, regardless of the value of $\mathrm{D_{HH}}$, the quantum discord exhibits pronounced extremal features when the normalized distance $r/r_h$ lies in the interval $\left[1.43, 1.5\right)$.

\section{conclusions}
In this work, based on the Werner state shared by Alice and Bob, we study the influence of the Hawking effect of the Dirac field in a Schwarzschild black hole on the quantum properties of the quantum atmosphere. Specifically, we analyze the evolution of quantum discord with the normalized distance, and find that quantum discord is redistributed between the physically accessible and inaccessible regions. Concretely, with the increase of the normalized distance, the quantum discord in the physically accessible region first decreases and then increases, while the variation trend of quantum discord in the physically inaccessible region is opposite. Remarkably, when the radial ratio $r/r_h$ lies in the interval $[1.43, 1.5)$, the quantum discord shows obvious extremum characteristics, which is closely related to the peak of Hawking radiation in the quantum atmosphere. By characterizing the dependence of quantum discord on the event horizon radius $r_h$ and the distance $r$ from the black hole center, we conclude that $D_{\text{HH}}$ can significantly enhance the distribution effect of quantum discord in both the physically accessible and physically inaccessible regions. Furthermore, the increase in $r_h$ will weaken the aforementioned distribution effect.
\begin{acknowledgments}
This work is supported by the Natural Science Foundation of Hainan Province under Grant No. 125RC744; the China Scholarship Council (CSC).
\end{acknowledgments}

\bibliography{ref}
\begin{figure*}
	\centering
	\includegraphics[width=1\textwidth, keepaspectratio]{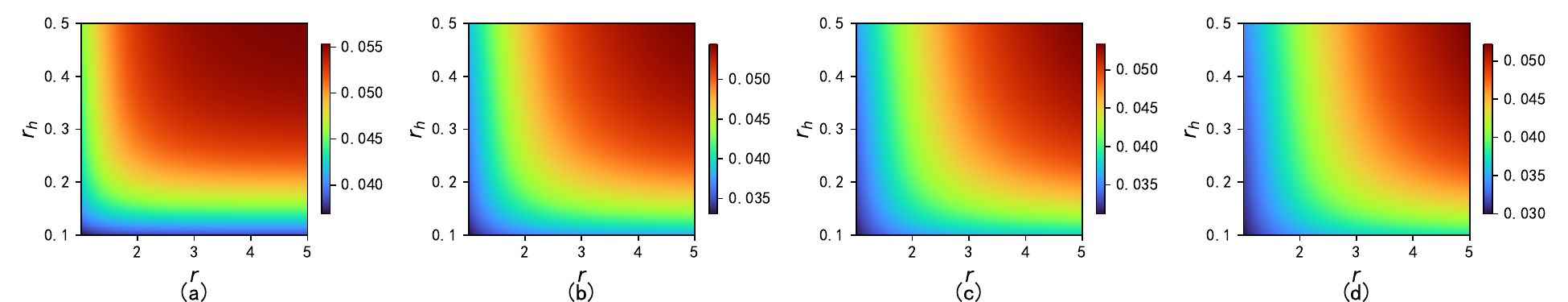}
	\caption{Physically accessible quantum discord D($\rho_{AB_\mathrm{I}}$) as a function of the distance from the black hole center ($r$) and the event horizon radius ($r_h$). Panel (a): $D_{\text{HH}} = 23.03$, panel (b): $D_{\text{HH}} = 50$, panel (c): $D_{\text{HH}} = 75$, panel (d): $D_{\text{HH}} = 100$. The parameter $p = 1$ is adopted for all panels.}
	\label{fig4}
	
	\vspace{8pt}
	\includegraphics[width=1\textwidth, keepaspectratio]{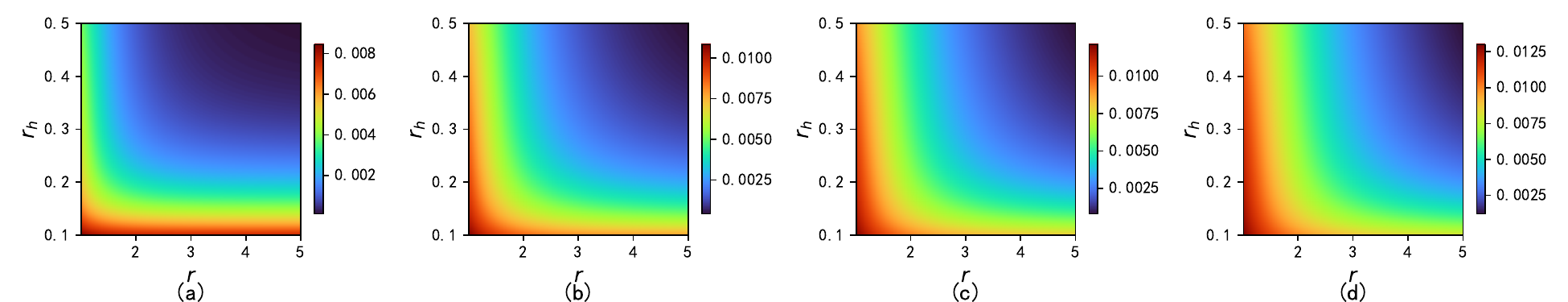}
	\caption{Physically inaccessible quantum discord $ D(\rho_{AB_\mathrm{II}}$) as a function of the distance from the black hole center $r$ and the event horizon radius $r_h$. Panel (a): $D_{\text{HH}} = 23.03$, panel (b): $D_{\text{HH}} = 50$, panel (c): $D_{\text{HH}} = 75$, panel (d): $D_{\text{HH}} = 100$. The parameter $p = 1$ is adopted for all panels.}
	\label{fig5}
\end{figure*}
\begin{figure*} 
	\centering
	\begin{minipage}[b]{0.48\textwidth}  
		\centering
		\includegraphics[width=\linewidth]{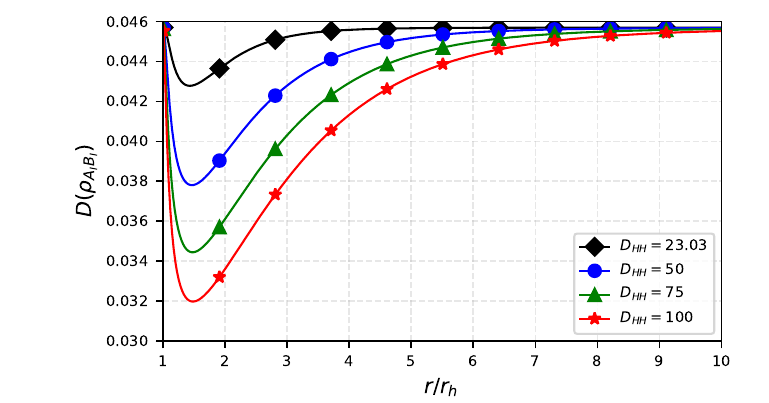}
		\\[-1.3em]
		(a)
	\end{minipage}
	\hfill 
	\begin{minipage}[b]{0.48\textwidth}
		\centering
		\includegraphics[width=\linewidth]{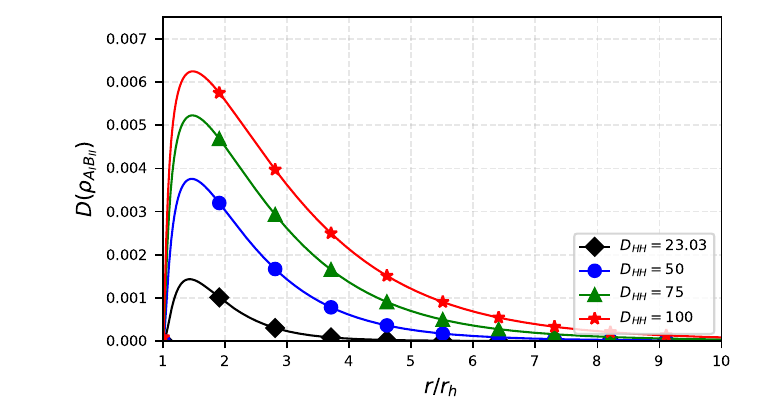}
		\\[-1.3em]
		(b)
	\end{minipage}
	
	\vspace{1.2em}
	\begin{minipage}[b]{0.48\textwidth}
		\centering
		\includegraphics[width=\linewidth]{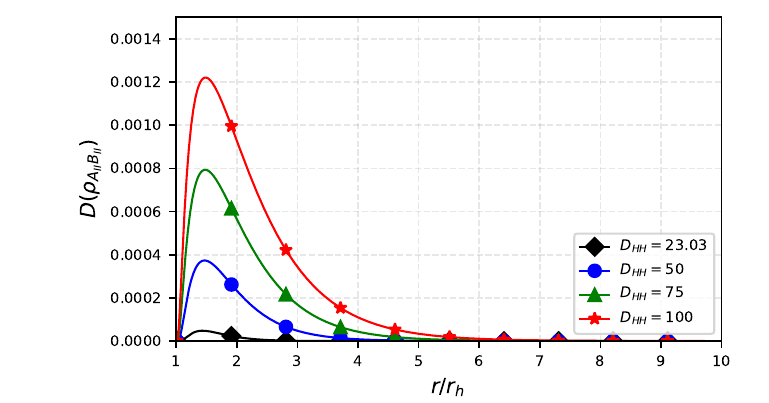}
		\\[-1.3em]
		(c)
	\end{minipage}
	\hfill
	\begin{minipage}[b]{0.48\textwidth}
		\centering
		\includegraphics[width=\linewidth]{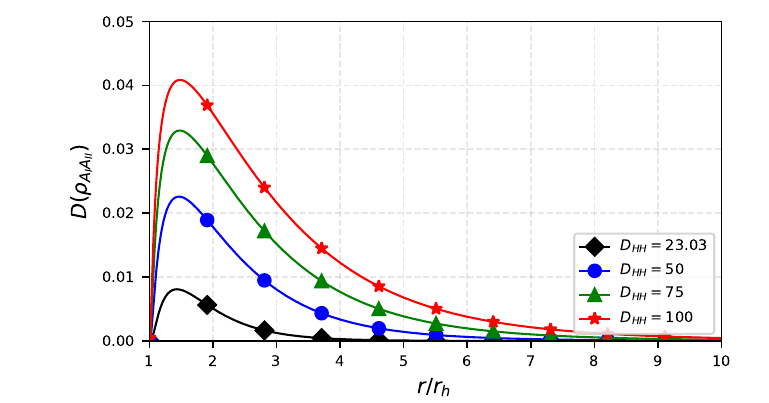}
		\\[-1.3em]
		(d)
	\end{minipage}
	
	\caption{The variation of the quantum discords $D(\rho_{A_\text{I} B_\text{I}})$, $D(\rho_{A_\text{I} B_\text{II}})$, $D(\rho_{A_\text{II} B_\text{II}})$ and $D(\rho_{A_\text{I} A_\text{II}})$ with the normalized distance $r/r_h$. The quantum discord is plotted as a function of $r/r_h$ for different values of the constant $\mathrm{D_{HH}}$ with the state parameter fixed at $p = \sqrt{10/11}$.}
	\label{fig6}
\end{figure*}
\appendix*
\section{}
It is worth noting that the linear map given by Eqs.~(8), exactly, induces a linear map on the operator space from $\mathbb{C}$ to $\mathbb{C} \otimes \mathbb{C}$ (here, we call it $\Lambda$). Specifically, the map $\Lambda$ acts on the unit matrices $\{|0\rangle\langle 0|, |0\rangle\langle 1|, |1\rangle\langle 0|, |1\rangle\langle 1|\}$, it has representations as follows,
\begin{align}
	\Lambda(|0\rangle\langle0|) &= \cos^2 \alpha |0\rangle\langle0|_\mathrm{I} \otimes |0\rangle\langle0|_\mathrm{II} \nonumber \\
	&\quad + \cos \alpha \sin \alpha |0\rangle\langle1|_\mathrm{I} \otimes |0\rangle\langle1|_\mathrm{II} \nonumber \\
	&\quad + \sin \alpha \cos \alpha |1\rangle\langle0|_\mathrm{I} \otimes |1\rangle\langle0|_\mathrm{II} \nonumber \\
	&\quad + \sin^2 \alpha |1\rangle\langle1|_\mathrm{I} \otimes |1\rangle\langle1|_\mathrm{II}, \\
	\Lambda(|0\rangle\langle1|) &= \cos \alpha |0\rangle\langle1|_\mathrm{I} \otimes |0\rangle\langle0|_\mathrm{II} \nonumber \\
	&\quad + \sin \alpha |1\rangle\langle1|_\mathrm{I} \otimes |1\rangle\langle0|_\mathrm{II}, \\
	\Lambda(|1\rangle\langle0|) &= \cos \alpha |1\rangle\langle0|_\mathrm{I} \otimes |0\rangle\langle0|_\mathrm{II} \nonumber \\
	&\quad + \sin \alpha |1\rangle\langle1|_\mathrm{I} \otimes |0\rangle\langle1|_\mathrm{II},  \\
	\Lambda(|1\rangle\langle1|) &= |1\rangle\langle1|_\mathrm{I} \otimes |0\rangle\langle0|_\mathrm{II}.
\end{align}
However, the Pauli matrices $\{\sigma_j\}$, where $j = 0, 1, 2, 3$ and $\sigma_0 = I$, form another well-known basis for the operator space $\mathbb{C}^{2\times 2}$. They have distinct representations under the map $\Lambda$, namely,
\begin{align}
	\Lambda(I) &= \cos^2\alpha\, |0\rangle\langle 0|_\mathrm{I} \otimes |0\rangle\langle 0|_\mathrm{II} \notag \\
	&+ \cos\alpha\sin\alpha\, |0\rangle\langle 1|_\mathrm{I} \otimes |0\rangle\langle 1|_\mathrm{II} \notag \\
	&+ \sin\alpha\cos\alpha\, |1\rangle\langle 0|_\mathrm{I} \otimes |1\rangle\langle 0|_\mathrm{II} \notag \\
	&+ \sin^2\alpha\, |1\rangle\langle 1|_\mathrm{I} \otimes |1\rangle\langle 1|_\mathrm{II} \notag \\
	&+ |1\rangle\langle 1|_\mathrm{I} \otimes |0\rangle\langle 0|_\mathrm{II}, \\
	\Lambda(\sigma_1) &= \cos\alpha\, (|0\rangle\langle 1|_\mathrm{I} + |1\rangle\langle 0|_\mathrm{I}) \otimes |0\rangle\langle 0|_\mathrm{II} \notag \\
	&+ \sin\alpha\, |1\rangle\langle 1|_\mathrm{I} \otimes (|0\rangle\langle 1|_\mathrm{II} + |1\rangle\langle 0|_\mathrm{II}), \\
	\Lambda(\sigma_2) &= \mathrm{i}\cos\alpha\, (|1\rangle\langle 0|_\mathrm{I} - |0\rangle\langle 1|_\mathrm{I}) \otimes |0\rangle\langle 0|_\mathrm{II} \notag \\
	&+ \mathrm{i}\sin\alpha\, |1\rangle\langle 1|_\mathrm{I} \otimes (|0\rangle\langle 1|_\mathrm{II} - |1\rangle\langle 0|_\mathrm{II}), \\
	\Lambda(\sigma_3) &= \cos^2\alpha\, |0\rangle\langle 0|_\mathrm{I} \otimes |0\rangle\langle 0|_\mathrm{II} \notag \\
	&+ \cos\alpha\sin\alpha\, |0\rangle\langle 1|_\mathrm{I} \otimes |0\rangle\langle 1|_\mathrm{II} \notag \\
	&+ \sin\alpha\cos\alpha\, |1\rangle\langle 0|_\mathrm{I} \otimes |1\rangle\langle 0|_\mathrm{II} \notag \\
	&+ \sin^2\alpha\, |1\rangle\langle 1|_\mathrm{I} \otimes |1\rangle\langle 1|_\mathrm{II} - |1\rangle\langle 1|_\mathrm{I} \otimes |0\rangle\langle 0|_\mathrm{II}.
\end{align}

Now, in order to obtain the subsystem, we trace out $\mathrm{II}$ mode, and we have the operators on $\mathrm{I}$ mode,
\begin{align}
	(I)_\mathrm{I} &= \cos^2\alpha\, |0\rangle\langle 0| + \sin^2\alpha\, |1\rangle\langle 1| \notag \\
	&+ |1\rangle\langle 1| = I - \sin^2\alpha\, \sigma_3, \\
	(\sigma_1)_\mathrm{I} &= \cos\alpha\, |0\rangle\langle 1| + \cos\alpha\, |1\rangle\langle 0| = \cos\alpha\, \sigma_1, \\
	(\sigma_2)_\mathrm{I} &= \mathrm{i}\cos\alpha\, |1\rangle\langle 0| - \mathrm{i}\cos\alpha\, |0\rangle\langle 1| = \cos\alpha\, \sigma_2, \\
	(\sigma_3)_\mathrm{I} &= \cos^2\alpha\, |0\rangle\langle 0| + \sin^2\alpha\, |1\rangle\langle 1| - |1\rangle\langle 1| = \cos^2\alpha\, \sigma_3.
\end{align}

Similarly, tracing out $\mathrm{I}$ mode, we have the operators on $\mathrm{II}$ mode
\begin{align}
	(I)_\mathrm{II} &= \cos^2\alpha\, |0\rangle\langle 0| + \sin^2\alpha\, |1\rangle\langle 1| + |0\rangle\langle 0|\notag \\
                    &= I + \cos^2\alpha\, \sigma_3, \\
	(\sigma_1)_\mathrm{II} &= \sin\alpha\, |0\rangle\langle 1| + \sin\alpha\, |1\rangle\langle 0| = \sin\alpha\, \sigma_1, \\
	(\sigma_2)_\mathrm{II} &= \mathrm{i}\sin\alpha\, |0\rangle\langle 1| - \mathrm{i}\sin\alpha\, |1\rangle\langle 0| = -\sin\alpha\, \sigma_2, \\
	(\sigma_3)_\mathrm{II} &= \cos^2\alpha\, |0\rangle\langle 0| + \sin^2\alpha\, |1\rangle\langle 1| - |0\rangle\langle 0| \notag \\
                            &= -\sin^2\alpha\, \sigma_3.
\end{align}

For any bipartite quantum state, it has Bloch representation like this,
\begin{equation}
	\begin{split}
		\rho = \frac{1}{4}\bigg(
		&I^A \otimes I^B
		+ \sum_{i} x_i \sigma_i \otimes I^B
		+ \sum_{j} y_j I^A \otimes \sigma_j \\
		&+ \sum_{i,j=1}^{3} g_{ij} \sigma_i \otimes \sigma_j
		\bigg)\textcolor{red}{.}
	\end{split}
\end{equation}
We assume that Alice remains in the asymptotically flat region, while Bob undergoes free fall into the Schwarzschild black hole. According to the transformation Eqs.~(8), the initial bipartite quantum state is converted into a tripartite quantum state. The partially traced reduced states can then be readily derived using  Eqs. (A.9) to (A.16).
\begin{equation}
	\begin{split}
		\rho_{AB_\mathrm{I}} &= \frac{1}{4}\bigg(
		I^A \otimes I^B 
		- I^A \otimes \sin^2\alpha\,\sigma_3 
		+ \frac{2p-1}{3}\cos\alpha\,\sigma_1 \otimes \sigma_1 \\
		&\quad + \frac{2p-1}{3}\cos\alpha\,\sigma_2 \otimes \sigma_2
		+ \frac{2p-1}{3}\cos^2\alpha\,\sigma_3 \otimes \sigma_3
		\bigg),
	\end{split}
\end{equation}

\begin{align}
	\begin{split}
		\rho_{AB_\mathrm{II}} &= \frac{1}{4}\bigg(
		I^A\otimes I^B + I^A\otimes \cos^2\alpha\,\sigma_3 +\frac{2p-1}{3}\sin\alpha\,\sigma_1\otimes\sigma_1 \\ 
		&\quad- \frac{2p-1}{3}\sin\alpha\,\sigma_2\otimes\sigma_2
		 - \frac{2p-1}{3}\sin^2\alpha\,\sigma_3\otimes\sigma_3
		\bigg).
	\end{split}
\end{align}

	We now consider the scenario where both Alice and Bob hover near the event horizon of a Schwarzschild black hole.
	Under the transformation given by Eq.~(8), the initial bipartite quantum state is mapped to a quadripartite quantum state.
	The partially traced reduced states can then be readily derived using  Eqs. (A.9) to (A.16).
\begin{equation}
	\begin{split}
		\rho_{A_\mathrm{I} B_\mathrm{I}}
		&= \frac{1}{4}\Bigl(
		I^A \otimes I^B
		- \sin^2\alpha \, \sigma_3 \otimes I^B
		- \sin^2\alpha \, I^A \otimes \sigma_3
		\\
		&\quad
		+ \frac{2p-1}{3}\cos^2\alpha \, \sigma_1 \otimes \sigma_1
		\\
		&\quad
		+ \frac{2p-1}{3}\cos^2\alpha \, \sigma_2 \otimes \sigma_2
		\\
		&\quad
		+ \Bigl[\sin^4\alpha + \frac{2p-1}{3}\cos^4\alpha\Bigr]
		\sigma_3 \otimes \sigma_3
		\Bigr),
	\end{split}
\end{equation}
\begin{equation}
	\begin{split}
		\rho_{A_\mathrm{I} B_\mathrm{II}}
		&= \frac{1}{4}\Bigl(
		I^A \otimes I^B
		- \sin^2\alpha \, \sigma_3 \otimes I^B
		+ \cos^2\alpha \, I^A \otimes \sigma_3
		\\
		&\quad
		+ \frac{2p-1}{3}\sin\alpha\cos\alpha \, \sigma_1 \otimes \sigma_1
		\\
		&\quad
		- \frac{2p-1}{3}\sin\alpha\cos\alpha \, \sigma_2 \otimes \sigma_2
		\\
		&\quad
		- \Bigl[\sin^2\alpha\cos^2\alpha + \frac{2p-1}{3}\sin^2\alpha\cos^2\alpha\Bigr]
		\sigma_3 \otimes \sigma_3
		\Bigr),
	\end{split}
\end{equation}
\begin{equation}
	\begin{split}
		\rho_{A_\mathrm{II} B_\mathrm{I}}
		&= \frac{1}{4}\Bigl(
		I^A \otimes I^B
		+ \cos^2\alpha \, \sigma_3 \otimes I^B
		- \sin^2\alpha \, I^A \otimes \sigma_3
		\\
		&\quad
		+ \frac{2p-1}{3}\sin\alpha\cos\alpha \, \sigma_1 \otimes \sigma_1
		\\
		&\quad
		- \frac{2p-1}{3}\sin\alpha\cos\alpha \, \sigma_2 \otimes \sigma_2
		\\
		&\quad
		- \Bigl[\sin^2\alpha\cos^2\alpha + \frac{2p-1}{3}\sin^2\alpha\cos^2\alpha\Bigr]
		\sigma_3 \otimes \sigma_3
		\Bigr),
	\end{split}
\end{equation}
\begin{equation}
		\begin{split}
			\rho_{A_\mathrm{II} B_\mathrm{II}}
			&= \frac{1}{4}\Bigl(
			I^A \otimes I^B
			+ \cos^2\alpha \, \sigma_3 \otimes I^B
			+ \cos^2\alpha \, I^A \otimes \sigma_3
			\\
			&\quad
			+ \frac{2p-1}{3}\sin^2\alpha \, \sigma_1 \otimes \sigma_1
			\\
			&\quad
			+ \frac{2p-1}{3}\sin^2\alpha \, \sigma_2 \otimes \sigma_2
			\\
			&\quad
			+ \Bigl[\cos^4\alpha + \frac{2p-1}{3}\sin^4\alpha\Bigr]
			\sigma_3 \otimes \sigma_3
			\Bigr),
		\end{split}
\end{equation}
\begin{equation}
	\begin{split}
		\rho_{A_\mathrm{I} A_\mathrm{II}}
		&= \frac{1}{4}\Bigl(
		I^A \otimes I^B
		- \sin^2\alpha \, \sigma_3 \otimes I^B
		+ \cos^2\alpha \, I^A \otimes \sigma_3
		\\
		&\quad
		+ \sin\alpha\cos\alpha \, \sigma_1 \otimes \sigma_1
		- \sin\alpha\cos\alpha \, \sigma_2 \otimes \sigma_2
		\Bigr).
	\end{split}
\end{equation}	

\end{document}